\documentstyle[12pt]{article}
\font\af=msbm12
 
\newcommand{\beeq}{\begin{equation}}
\newcommand{\eneq}{\end{equation}}
\newcommand{\beqn}{\begin{eqnarray}}
\newcommand{\eeqn}{\end{eqnarray}}
\def\di{D\!\!\!\!\slash_A\,}

\def\Z{\mbox{\af Z}}
\def\ra{\rangle}
\def\la{\langle}
\begin{document}
\title{ Witten-Veneziano Relation for the Schwinger Model}

\author{S.Azakov $^{1}$\footnote{e-mail address: azhep@lan.ab.az}~,
H.Joos $^{2}$\footnote{e-mail address: joos@x4u2.desy.de}~
and A.Wipf $^{3}$ \footnote{e-mail address: wipf@tpi.uni-jena.de}}
\maketitle

\noindent
$^{1}${{\it Inst. of Physics, Azerbaijan Academy of Sciences,
H.Cavid ave.33, Baku 143, Azerbaijan}\\
$^{2}${{\it DESY, Notkestrasse 85, D-22603 Hamburg, Germany}}\\
$^{3}${{\it Theor.-Phys. Institut, Universit{\"a}t Jena, Fr{\"o}belstieg
1, D-07743 Jena, Germany}}\\
\begin{abstract}
The Witten-Veneziano relation between 
the topological susceptibility of pure gauge
theories without fermions and the main contribution 
of the complete theory and the corresponding formula 
of Seiler and Stamatescu with the so-called contact
term are discussed for the Schwinger model on a circle. 
Using the (Euclidean) path  integral and the 
canonical (Hamiltonian) approaches at finite temperatures
we demonstrate that both formulae give the same result in 
the limit of infinite volume and (or) zero temperature.
\end{abstract}

The Witten--Veneziano relation (WVR) \cite{Wit,Ven}
plays an important role in identifying
topological excitations in gauge theories with fermions. It relates
the topological susceptibility of pure gauge theory without fermions
to the main fermionic contributions of the complete theory. Since in the
general derivation some approximations are involved, it is by educational
reasons worthwhile to study models in which the WVR is exact. We choose
the Schwinger Model (SM) \cite{SS,Gat}. However, a correct treatment
requires
to start with the SM on a compact space, and to consider the infinite
space-time
limit only at the end.
\par
According to Witten  and Veneziano for the SM the following formula
\beeq
m_{\eta'}^2=\frac{4}{|f_1|^2}\chi^{(0)}_{top}
\label{m1}
\eneq
should hold. Here $m_{\eta'}$ is a mass of the $U(1)$ pseudoscalar
$"\eta'$ meson", $f_1$ is its decay constant, and $ \chi^{(0)}_{top}$ is a
topological susceptibility in pure $U(1)$ theory. For the SM we have
$m_{\eta'}= m=\frac{e}{\sqrt{\pi}}$ and $f_1 =\frac{i}{\sqrt{\pi}}$, 
i.e. the "massive meson" of the SM is identified with $\eta'$ particle.
\par
Seiler and Stamatescu \cite{SS} reanalyzed the WVR
in Euclidean space-time and their final result, applied to the SM, 
reads
\beeq
m_{\eta'}^2=\frac{4}{|f_1|^2}P^{(0)}(0)~~.
\label{m2}
\eneq
The topological susceptibility of pure gauge theory in the WVR
has been replaced by the contact term $P^{(0)}(0)$ of the two point
function of the topological charge density
in the theory with massless fermions. In
their
work they calculated $P^{(0)}(0)$ and showed that formula (\ref{m2})
really holds.
\par
We calculate in the present work $\chi^{(0)}_{top}$  and $P^{(0)}(0)$
for the SM on the Euclidean torus, and $\chi^{(0)}_{top}(0)$
on a  circle at finite temperature. Finally, we show that at
finite temperature ~(\ref{m1}) and ~(\ref{m2}) are correct and
that they agree in the
limit of infinite volume and
(or) zero temperature.

\section {Euclidean path integral approach on the torus}
The topological susceptibility $\chi_{top}$ in the theory on the
two-dimensional Euclidean torus ${\cal T}$ (with circumferences
$L_1,L_2$ and volume $V=L_1L_2$) is defined as follows 
\beeq
\chi_{top}=\int_{\cal T} \la q(x)q(0) \ra \,d^2x ~,
\label{topsus1}
\eneq
where $q(x)$ is the topological charge density
\beeq
q(x)=\frac{e}{2\pi}F_{12}(x)\equiv \frac{e}{2\pi}E(x)
\label{tcd}
\eneq
and
\beeq
\int q(x)\,d^2x =\frac{e}{2\pi}\int E(x)\,d^2x =k ~,
~~~~~k\in \Z
\label{topchar}
\eneq
is the integer-valued topological charge or instanton number
in the two-dimensional $U(1)$ gauge
theory, $E(x)=\partial_1 A_2(x) -\partial_2 A_1(x) $ is the field
strength. Here and below the spacetime-integrals extend over the
torus ${\cal T}$. From Eqs (\ref{topsus1}) and (\ref{tcd}) we obtain
\beqn
\chi_{top}= \frac{e^2}{(2\pi)^2}
\int \la E(x)E(0) \ra \,d^2x ~~.
\label{topsus2}
\eeqn
The expectation values $\la O(A)\ra $ of an operator $O(A)$ which depends
only
on the gauge field $A_{\mu}(x)$ are different in the theory with fermions,
e.g. the SM, and in the pure electrodynamics on the torus.
After integration with respect to fermions in the SM we have 
\cite{SW,JA,SW1,Aza}
\beeq
\la O(A) \ra =\frac{1}{Z}\int_{{\cal A}_0} {\cal D}A \det \di O(A)
e^{-\frac{1}{2}\int d^2x E^2(x)}~,
\label{qev}
\eneq
where $\di$ is a Dirac operator in the external electromagnetic field.
As a consequence of the index theorem, there are zero modes in the 
sectors with instanton number $k \neq 0 $ \cite{Jo}. Therefore
the renormalized $\det \di$ is unequal to zero only in the trivial
sector ${\cal A}_0$ of the gauge field configurations with
vanishing topological charge. $Z$ is the partition function in this
theory 
\beeq
Z=\int_{{\cal A}_0} {\cal D}A \det \di
e^{-\frac{1}{2}\int d^2x E^2(x)} ~~.
\label{pf}
\eneq
In pure gauge theory (where expectation values will be denoted by $\la
\cdots \ra_0$) all instanton sectors
contribute to the expectation value $\la O(A) \ra_0$:
\beeq
\la O(A) \ra_0 =\frac{1}{Z_g}\sum_{k\in \Z}\int_{{\cal A}_k} {\cal D}A\, O(A)
e^{-\frac{1}{2}\int d^2x E^2(x)}
\label{qev0}
\eneq
and the partition function reads
\beeq
Z_g =\sum_{k\in Z}\int_{{\cal A}_k} {\cal D}A
e^{-\frac{1}{2}\int d^2x E^2(x)}~~.
\label{pf0}
\eneq
In both theories in the sector ${\cal A}_k$ with topological charge
$k$ the gauge potential has the form
\beeq
A_{\mu}^{(k)}(x) = A_{\mu}^{(0)}(x)+C_{\mu}^{(k)}(x)~~,
\eneq
where $ A_{\mu}^{(0)}(x)$ is a single valued "continuous" function on
${\cal T}$ and $C_{\mu}^{(k)}(x)$ is a global
instanton-type potential which in the Lorentz gauge reads
\beeq
C_{\mu}^{(k)}(x)=- \frac{\pi k}{e V}\epsilon_{\mu\nu}x_{\nu}~.
\eneq
For $ A_{\mu}^{(0)}(x)$ we may use the Hodge decomposition
\beeq
A_{\mu}^{(0)}(x)=
\partial_\mu a(x)+t_{\mu}+\epsilon_{\mu\nu}\partial_{\nu}b(x)~~,
\label{hdec}
\eneq
where $\partial_\mu a(x)$ is pure gauge, $t_{\mu}$ is a (constant) toron
field restricted  to the dual torus $ 0\leq t_{\mu}\leq
T_{\mu}\equiv 2\pi/eL_{\mu}$, $\epsilon_{\mu\nu}\partial_{\nu}b(x)$
is a curl and $a(x)$ and $b(x)$ are continuous on ${\cal T}$ and
orthogonal to the constant functions:
$\int a(x)\, d^2x=\int b(x)\,d^2x=0$ (on the torus the
Laplacian  $\Box \equiv \partial_1^2+\partial_2^2$ is invertible
only
on
functions which integrate to zero).
\par
So the path measure in Eqs.(\ref{qev}) - (\ref{pf0}) has a form
\beeq
\int{\cal D}A\cdots = \int{\cal D}b\int{\cal D}a
\int_0^{T_1}dt_1\int_0^{T_2}dt_2 \cdots~~.
\eneq
The two-point function $\la E(x)E(y) \ra $ has been calculated
in the SM on the torus \cite{SW,Aza} with the following result:
\beeq
\la E(x)E(y) \ra = \delta(x-y) -m^2 G_m(x-y)~~,
\label{ff}
\eneq
where $\delta (x)$ is the $\delta $-function on the torus and
\beeq
G_m(x)= \frac{1}{V}\sum_{n_1,n_2}
\frac{e^{2\pi i\left(n_1 x_1/L_1+n_2x_2/L_2\right)}}
{m^2 +\left(\frac{2\pi}{L_1}\right)^2n_1^2 +\left(\frac{2\pi}{L_2}\right)
^2n_2^2}
\eneq
is the Greens function of massive scalars on it.
From Eqs.(\ref{topsus2}) and (\ref{ff}) we see that the contact term is
\begin{equation} \label{cnt22}
P^{(0)}(0)  = \frac{e^2}{4\pi^2}.
\end{equation}
As is generally true in gauge theory with massless fermions, 
the topological susceptibility $\chi_{top}$ vanishes in the 
SM and therefore the relation
(\ref{m2})
holds.
\par
Now let us calculate the two-point function $\la E(x)E(y)\ra_0$
in pure electrodynamics.
Using the decomposition Eq.(\ref{hdec}) we get for the field strength
\beeq
E(x) = - \Box b(x)+\frac{2\pi k}{eV}
\eneq
and for the action
\beeq
\frac{1}{2}\int d^2x E^2(x) = \pi\tau k^2
+\frac{1}{2}\int d^2x\, b(x)\Box^2 b(x),\qquad
\tau=2\pi/e^2V.
\eneq
Then
\beqn
\la E(x)E(y)\ra_0 &=&\left(\frac{2\pi }{eV}\right)^2
\frac{\sum k^2e^{-\pi\tau k^2}}
{\sum e^{-\pi\tau k^2}} 
+ \frac{\int {\cal D}b\, \Box b(x)\Box b(y)e^{-\frac{1}{2}\int d^2x\,
b(x)\Box^2 b(x)}}{{\int {\cal D}b\, e^{-\frac{1}{2}\int d^2x\,
b(x)\Box^2 b(x)}}} \nonumber \\
&=&\left(\frac{2\pi }{eV}\right)^2
\frac{\sum k^2e^{-\pi\tau k^2}}
{\sum e^{-\pi\tau k^2}} +\delta (x-y)
-\frac{1}{V}~, \label{ff0}
\eeqn
where one sums over all $k\in \Z$.
The presence of the last term in Eq.(\ref{ff0}) is due to the fact that
$b(x)$ does not have a zero mode since it integrates to zero.
From Eqs(\ref{topsus2}) and (\ref{ff0}) we get for the topological
susceptibility in pure electrodynamics
\beqn
\chi_{top}^{(0)} = \frac{1}{V}\,\frac{\sum k^2e^{-\pi\tau
k^2}}{\sum e^{-\pi\tau k^2}}~~.
\label{topsus01}
\eeqn
In pure electrodynamics all instanton sectors contribute to
the topological susceptibility. This remains true 
in infinite volume limit, as we shall see below.
\par
Using the definition of the Jacobi's $\theta_3$ function \cite{Bat}
\beqn
\theta_3(z|\tau) =\sum_{n=-\infty}^{n=\infty} e^{-\tau\pi n^2 +2\pi inz}
\eeqn
we may rewrite Eq.(\ref{topsus01}) in their terms:
\beqn
\chi_{top}^{(0)} = -\frac{1}{4
\pi^2V}\frac{\theta_3^{\prime\prime}(0|\tau)}
{\theta_3(0|\tau)}~~.
\label{topsus02}
\eeqn
\section{Topological susceptibility  in pure electrodynamics on a circle}
Pure electrodynamics in two dimensions is defined in a non-trivial way
only on a
compact space where it has non-trivial gauge invariant solutions.
Therefore we consider it again on a circle with circumference $L_1$.
Manton \cite{Man} was the first who considered pure electrodynamics
on a circle and showed that in
this model there is not a unique canonical quantization, because the
representation of the electric field operator contains an arbitrary real
parameter $\theta$. The Hamiltonian has eigenvalues: $E_k
=\frac{1}{2} L_1e^2\left(k +\theta/2\pi \right)^2,~ k\in \Z$.
In this theory
the $\theta$ angle (the fractional part of $\theta/2\pi$) 
is a relevant parameter and different values of $\theta$
separate different worlds.
It corresponds to the famous $\theta$ angle of $SU(n)$ gauge theories.
In contrast, as in the case of quantum chromodynamics with massless fermions
\cite{Ja},
in the SM the angle $\theta$ plays non physical role.
\par
The topological susceptibility in this case reads
\beqn
\chi_{top}^{(0)}(0) = \frac{1}{L_1}\frac{\partial^2 F(\theta)}
{\partial \theta^2}|_{\theta=0} ~,
\label{topsusm1}
\eeqn
where $ F(\theta) =
-\frac{1}{\beta}\log Z(\theta)$ 
is the free energy and $Z(\theta)$ the partition function
at temperature $T=1/\beta$:
\beqn
Z(\theta) = \sum_k e^{-\beta E_k(\theta)} =
\sum_k e^{-\frac{1}{2}\beta L_1 e^2( k+\theta/2\pi)^2}~~.
\label{pfm}
\eeqn
From Eqs.(\ref{topsusm1}) and (\ref{pfm}) it follows that
\beqn
\chi_{top}^{(0)}(0)&=& - \frac{1}{L_1
\beta}\frac{Z^{\prime\prime}(0)}
{Z(0)}
=\frac{e^2}{4\pi ^2}\Bigg(1 - \beta L_1e^2\,
\frac {\sum k^2 e^{-\frac{1}{2}\beta L_1 e^2 k^2 }}
{\sum e^{-\frac{1}{2}\beta L_1 e^2k^2}} \Bigg).
\eeqn
Now we can use the following transformation formula \cite{Bat} between
$\theta_3$-functions of zero
argument
\beqn
\theta_3(0|\tau)
=\frac{1}{\tau}\;\theta_3\Big(0|-1/\tau\Big)~~,
\eeqn
and prove that
\beqn
\chi_{top}^{(0)}(0)=\chi_{top}^{(0)}~~,
\eeqn
if we take $\beta=L_2$.
Thus the path integral approach and canonical approach give the same result
for the topological susceptibility.
\par
A systematic comparison between the Hamiltonian
approach for the SM on a circle and the Euclidean path integral
approach on the torus was done in the forthcoming paper
\cite {AJ}. There it is shown how to obtain Eq.(\ref {ff})
within the Hamiltonian approach.

\section{The infinite volume limit}
In order to consider the limits of infinite volume ($L_1\rightarrow
\infty$)
and/or the zero temperature $(L_2\rightarrow \infty)$ we will use the
following expansion \cite{Tolke}:
\beqn
\frac{\theta_3^{\prime\prime}(0|\tau)}
{\theta_3(0|\tau)}= -\frac{2\pi}{\tau}
-\frac{8\pi^2}{\tau^2}e^{-\pi/\tau}
\big(1-2\,e^{-\pi/\tau}\big) +\cdots
\eeqn
Then from Eq.(\ref{topsus02}) we find
\beqn
\chi_{top}^{(0)} =\frac{e^2}{4\pi^2} +\cdots~~,
\eeqn
where $\cdots$ are terms which disappear if at least one of 
circumferences $L_i$ tends to infinity. Thus we have 
shown that in this cases the topological
susceptibility (\ref{topsus02}) agrees with the contact 
term (\ref{cnt22}).

\noindent
{\bf Acknowledgment:}
S.A. would like to thank DAAD for a
grant which allowed to realize this project.


\begin{thebibliography}{99}
\bibitem{Wit}
E.Witten, Nucl.Phys. {\bf B149} (1979) 285.
\bibitem{Ven}
G.Veneziano, Nucl.Phys. {\bf B159} (1979) 213.
\bibitem{SS}
E.Seiler and I.O.Stamatescu, preprint MPI-PAE/PTh 10/87, 1987.
\bibitem{Gat}
C.Gattringer, Ann.Phys. {\bf 250} (1996) 389; Ph.D.thesis,
MPI-Ph/95-24, hep-th 9503137.
\bibitem{SW}
I.Sachs and A.Wipf, Helv.Phys.Acta {\bf 65} (1992) 652.
\bibitem{JA}
H.Joos and S.Azakov, Helv.Phys.Acta {\bf 67} (1994) 723.
\bibitem{SW1} I.Sachs and A.Wipf, Phys.Lett. {\bf B326} (1994) 105.
\bibitem{Aza}
S.Azakov, Fortschr.Phys. {\bf 45} (1997) 589.
\bibitem{Jo} H.Joos, Helv.Phys.Acta {\bf 63} (1990) 652.
\bibitem{Man}
N.Manton, Ann.Phys.(N.Y.) {\bf 159} (1985) 220.
\bibitem{Ja} R.Jackiw and C.Rebbi, Phys.Rev.Lett. {\bf 37} (1976) 172.
\bibitem{Bat}
A.Erd{\'e}lyi (Director), {\it Higher Transcedental Functions},
Vol.2, Chapter 13.
\bibitem{AJ}
S.Azakov and H.Joos, in preparation.
\bibitem{Tolke}
F.T{\"o}lke, {\it Praktische Funktionenlehre}, Zweiter Band,
Springer-Verlag, 1966.
\end{thebibliography}
\end{document}